\documentclass{article}
\usepackage[backend=biber]{biblatex}
\usepackage{amsmath}
\usepackage{amssymb}
\usepackage{graphicx}
\usepackage{subcaption}
\usepackage{float}
\usepackage{setspace}
\title{Payrolls to Prompts: Firm-Level Evidence on the Substitution of Labor for AI}
\author{Ryan Stevens\thanks{Director of Applied Sciences, Ramp}}
\date{January 2026}

\begin{document}
\onehalfspacing

\maketitle
\newpage

\section{Introduction}
AI is profoundly reshaping the nature of work. Every day, stories abound about startups disrupting industries, new models being released, and the anxiety of legacy companies getting left behind in the new AI era. An important question in this sector is whether firms are substituting labor for AI. Given its potential implications for labor markets, productivity, and income inequality, the magnitude and type of substitution is critical to understand. While this question is often asked, there is limited evidence on how businesses are spending on AI, and how that is interacting with the labor force. This study provides some of the first direct evidence on how firms are spending on AI tools, and how that is interacting with the labor force. We focus on two key questions: Are businesses substituting labor for AI? If so, what is the rate of substitution between AI and labor? Put differently, what are the cost savings from AI adoption?

This study focuses on a specific sector of workers most impacted by AI: jobs contracted through online labor marketplaces. Online labor marketplaces are a natural choice for studying the early effects of AI adoption because the types of tasks being performed are well-suited for automation. We leverage a unique firm-level dataset that contains spending data from thousands of companies. For each firm, we are able to track their spending on online labor marketplaces and AI model providers from 2021 to 2025. This gives us a unique ability to directly observe firm-level substitution patterns and understand the heterogeneity in substitution across firms.

We find that firms are replacing labor with AI. These effects are not uniform across firms, we find heterogeneity in the rate of substitution by firm exposure to AI shocks. Specifically, businesses that spent higher shares of their total spend on online marketplaces were more likely to substitute labor for AI. Additionally, AI appears to be a cost-saving technology; businesses are not substituting \$1 of labor for \$1 in AI. Instead, we find that businesses most exposed to AI shocks substitute labor for AI at a rate of \$1 for \$0.03 in AI spending. This implies a 20 to 25x cost savings from AI adoption.

While we find that firms are substituting labor for AI, this does not imply that in aggregate, labor will be replaced by AI. The evidence we present could be consistent with a story where AI is labor-augmenting, and aggregate labor demand increases. For example, demand for workers that can deploy and maintain AI systems increases at a rate faster than the rate at which labor is being replaced. What we see in this paper are micro-level substitution patterns not aggregate labor market outcomes.

\section{Related Literature}
There is a growing literature (both in popular media and academic research) on the impact of AI and jobs.

\textbf{Measuring automation and AI exposure.} Building on task-based models of technological change \cite{AutorLevyMurnane2003, AcemogluAutor2011}, a foundational body of work develops quantitative measures of which occupations and tasks are most susceptible to automation by digital technologies. \cite{FreyOsborne2017} pioneer this approach by estimating that 47\% of U.S.\ employment is at high risk of computerisation. \cite{BrynjolfssonMitchell2017} and \cite{BrynjolfssonMitchellRock2018} develop a framework for assessing which tasks current machine-learning systems can perform, emphasizing that AI’s impact depends on the specific capabilities required by each task rather than at the occupation level. \cite{Webb2020} constructs measures of AI exposure by linking patent text to occupational task descriptions. \cite{FeltenRajSeamans2019, FeltenRajSeamans2020, FeltenRajSeamans2021, FeltenRajSeamans2021b} develop ability-based occupational exposure indices that link AI capabilities to occupational abilities, providing widely used measures of which jobs are most exposed to AI advances. More recently, \cite{FeltenRajSeamans2023} adapt this approach to generative AI, and \cite{EloundouManningMishkinRock2024} build GPT-4-based exposure indices at the task and occupation level.

\textbf{Compositional effects of AI on labor using wage data.}
A growing literature combines these exposure measures with administrative or survey data to study realized labor-market adjustments. \cite{EloundouManningMishkinRock2024} construct task-based exposure indices for GPT models and show that potential impacts are concentrated in higher-wage, higher-skill occupations, with substantial heterogeneity across tasks and industries. \cite{AcemogluAutorHazellRestrepo2022} link AI-related vacancies to detailed occupation data and document that AI adoption reshapes the composition of labor demand rather than uniformly reducing employment. \cite{JohnstonMakridis2025} follow up on this work using a difference-in-differences design and granular administrative data from the U.S.\ labor market, exploiting variation in large language model exposure across state–industry cells; they find that sectors more exposed to LLMs tend to experience relative gains in employment and wages, particularly for younger and more educated workers, while aggregate effects on wages are more muted. Building on similar exposure measures but using high-frequency payroll data, \cite{BrynjolfssonChandarChen2025} document that, following the diffusion of generative AI tools, employment for early-career workers in highly AI-exposed occupations declines sharply relative to older workers, whereas employment for more experienced workers is broadly stable and wage adjustments are limited. Together, these studies suggest that generative AI induces heterogeneous, compositionally important changes in labor demand across age, skill, and exposure groups rather than a uniform decline in employment.

\textbf{Studies of AI on online labor marketplaces.}
Online freelancing platforms offer a useful setting to study generative AI's short-run labor-market impact. \cite{DemirciHannaneZhu2025} find large declines in job postings for AI-exposed tasks such as writing, coding, and image creation. \cite{HuiReshefZhou2024} show that freelancers in AI-exposed occupations experience reductions in both employment and earnings. \cite{TeutloffEtAl2025} find that demand for substitutable skills (writing, translation) falls by 20-50\%, while complementary skills (ML programming, chatbot development) see stable or increased demand. \cite{LiuXuNanLiTan2025} document market contraction in LLM-aligned submarkets and show that ChatGPT enables skill transitions into programming tasks. These studies focus on worker-level outcomes and job postings but do not observe the firms posting jobs. Our analysis complements this literature by studying firm-level substitution patterns directly.

\section{Dataset}
This study uses data from Ramp, an expense management platform built to help businesses manage their spending. Ramp has both corporate card and bill invoice product lines. Business to business payment happens via two primary payment methods: cards and bank-to-bank transfers. In the United States, bank-to-bank transfers usually happen via automated clearing house (ACH) transfers. Bank-to-bank transfers are the dominant payment method for businesses. In our dataset, 66\% of all spending happens via bank-to-bank transfers. Thus, we are able to observe spending on a wide range of goods and services, including online labor marketplaces and AI model providers.

Each bill invoice and card transaction is matched, where possible, to a merchant using Ramp’s internal merchant database. Payment methods don’t provide a standardized, comprehensive merchant database. For example, card networks include merchant category codes (MCCs), but these alone are insufficient for consistent merchant identification. Companies develop their own merchant databases, based on information provided by payment processors, as well as, customer-provided information. We leverage Ramp's proprietary merchant database to identify the merchant for each bill invoice and card transaction.

Our dataset contains spending data from Q3 2021 to Q3 2025. We include firms that were spending at least \$2,500 in at least two consecutive months in 2021 to prevent including companies that enter later or ramp up their spending later in the sample period. For our regression analysis, we exclude firms that were spending less than \$2,500 in Q3 2025 to prevent firm exit from impacting our substitution analysis. 

To focus on AI adoption, we include periods prior to the introduction of ChatGPT in October 2022. This event marks a natural experiment that led to a sharp increase in spending on AI model providers. We will discuss our identification strategy in more detail later in the paper.

We focus on a specific marketplace: online labor marketplaces and AI model providers. We define online labor marketplaces as any of the following merchants: Upwork, Fiverr, Toptal, PeoplePerHour, Arc (formerly Codementor), MarketerHire, and Catalant. We define AI model providers as any of the following merchants: OpenAI and Anthropic. We do not include other AI model providers such as Google, Meta, or Microsoft in this analysis due to difficulty uniquely identifying spend on their products. Given our focus on online labor marketplaces, we only include companies that had positive spending on online labor marketplaces in both Q1 2022 and Q2 2022, i.e., the quarters before the introduction of ChatGPT.

\section{Aggregate Spending Patterns}

Before we dive into firm-level substitution, we first look at aggregate spending patterns to motivate our identification strategy.

We first look at the aggregate total share of spending on online labor marketplaces and AI model providers. We find that the share of spending on online labor marketplaces has decreased significantly, from 0.66\% in Q4 2021 to 0.14\% in Q3 2025. On the other hand, the share of spending on AI model providers has increased significantly, from 0\% in Q4 2021 to 2.85\% in Q3 2025. 

\begin{figure}[ht]
\centering
\includegraphics[width=0.8\textwidth]{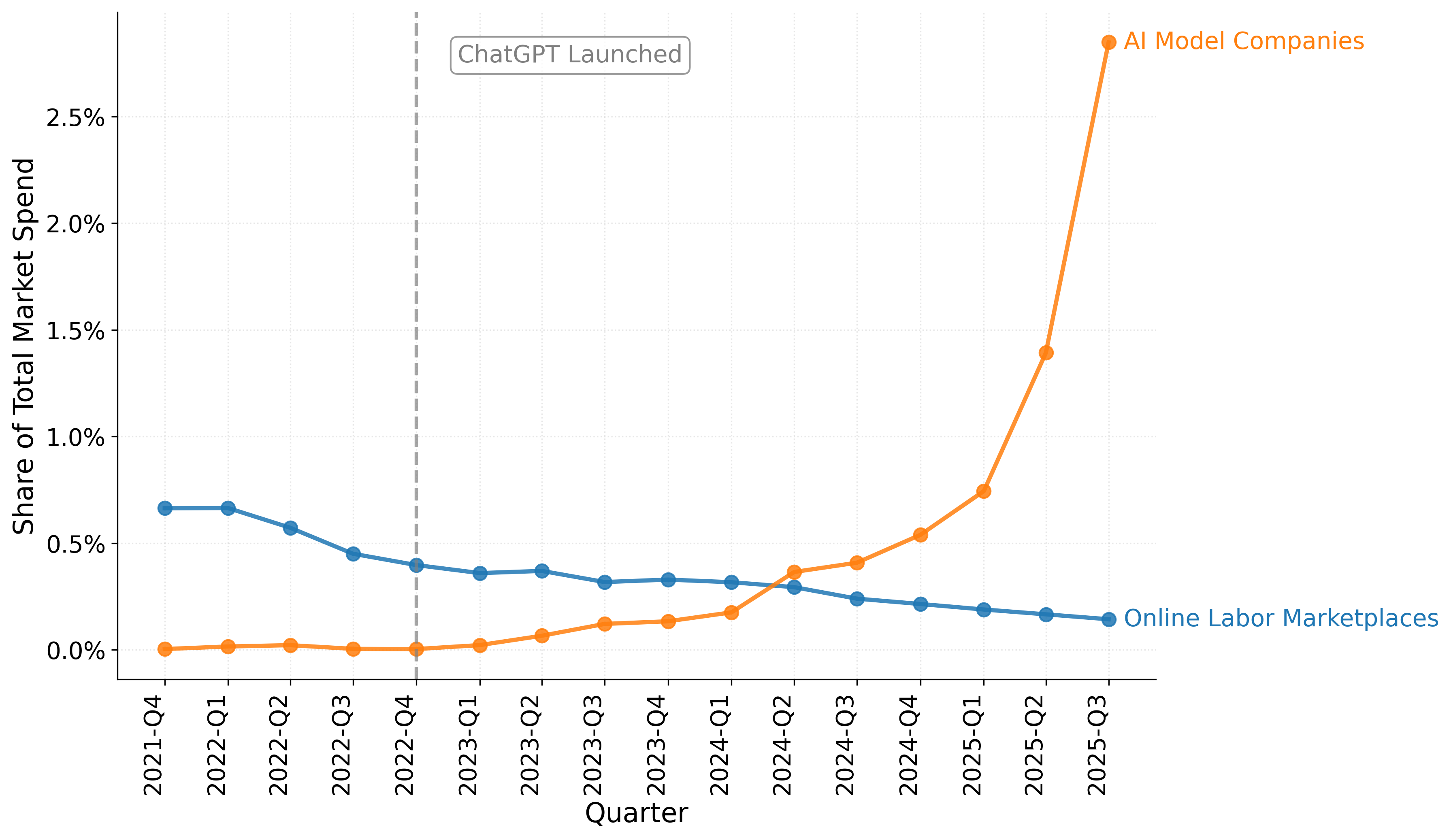}
\caption{Share of total spend on online labor marketplaces and AI model providers over time}
\label{fig:share_over_time}
\end{figure}

Firms may still be increasing investment in labor marketplaces even as the share of spending falls. This could happen if they are spending on other goods and services. We find this is not the case. Aggregate quarterly growth in online labor marketplace spending turned negative by Q3 2024 and has continued to remain negative. In contrast, AI model provider spending has seen positive growth since Q3 2022.

\begin{figure}[ht]
\centering
\includegraphics[width=0.8\textwidth]{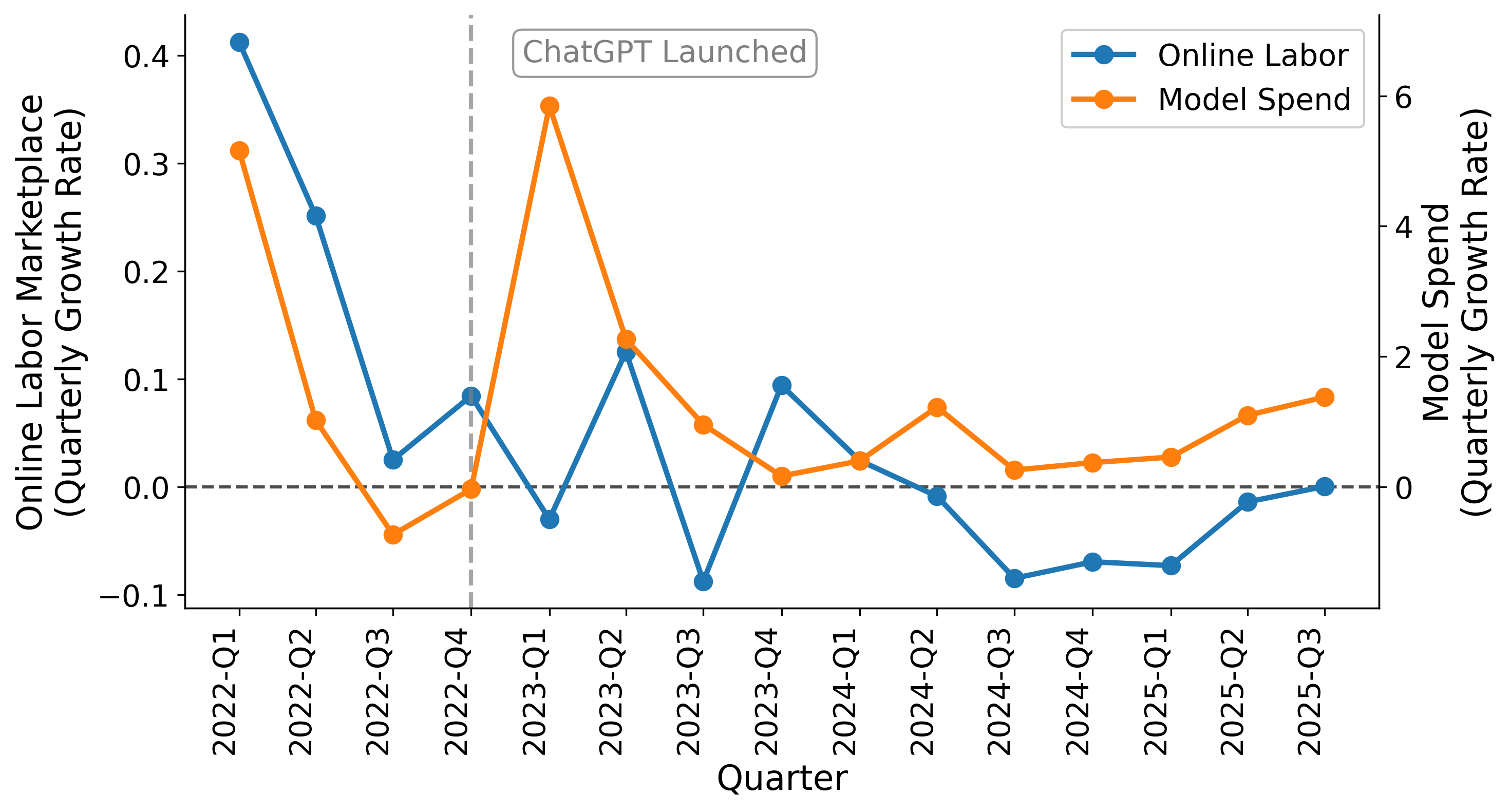}
\caption{Quarterly growth rates for online labor marketplace and AI model provider spending}
\label{fig:quarterly_growth}
\end{figure}

The prior figures focus on aggregate market-wide spending patterns. This does not tell us whether individual firms are substituting labor for AI. Our aggregate growth patterns suggest there are structural shifts from labor to AI, but we want to understand if this is a firm-level effect or a compositional effect, i.e., are the types of firms changing, or are the spending patterns within firms changing?

Our dataset allows us to answer this question. For each firm, we calculate the share of spend on online labor marketplaces and AI model providers in Q2 2022 (the quarter before the introduction of ChatGPT) and Q2 2025. We compare the same quarters across 2022 and 2025 to control for seasonality in business spending patterns. To do so, we plot binned probability distribution functions of quarterly spending shares. 

We observe a noticeable shift from labor to AI. More than 50\% of businesses that had spent on online labor marketplaces in Q2 2022 spent 0\% in Q2 2025, whereas roughly 80\% of businesses spent between 0 and 5\% of their total spend on AI model providers in Q2 2025. The effect is clear and dramatic.

\begin{figure}[ht]
\centering
\includegraphics[width=0.48\textwidth]{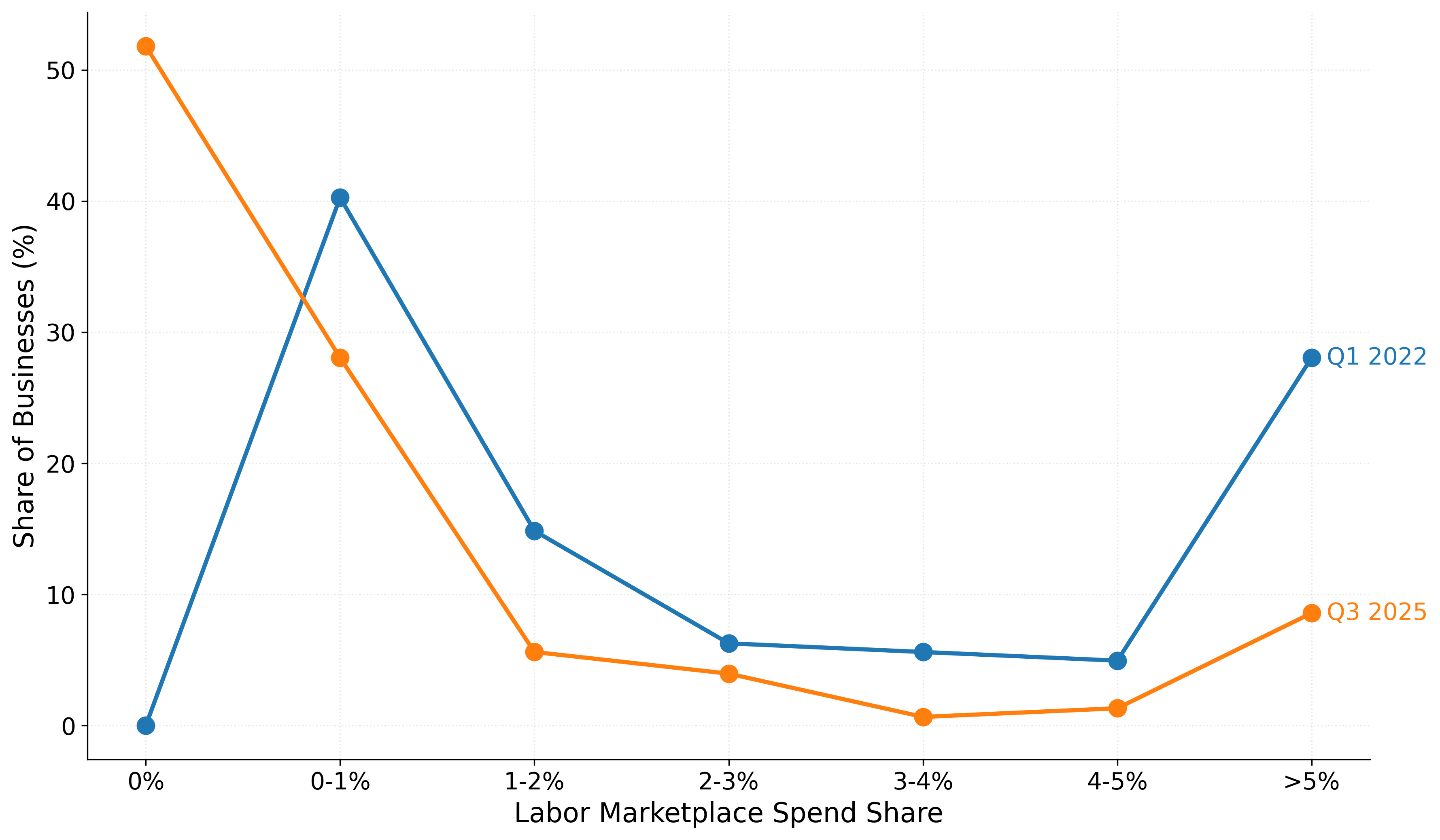}
\includegraphics[width=0.48\textwidth]{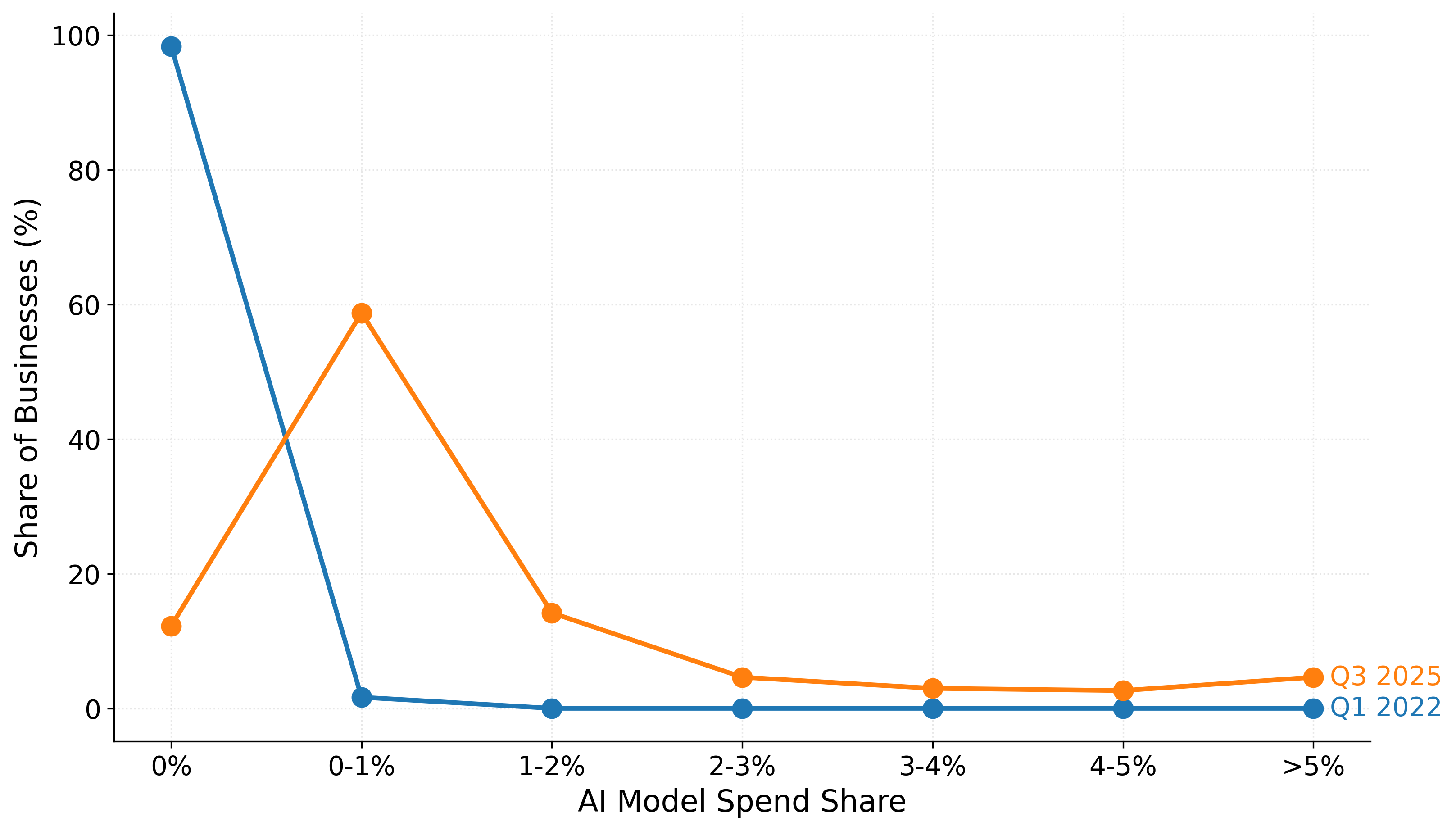}
\caption{Distribution of spending share on online labor marketplaces (left) and AI model providers (right) comparing Q2 2022 to Q2 2025}
\label{fig:spend_distributions}
\end{figure}

None of the evidence is causal; it could be that compositional effects are at play. Some firms may be decreasing investment in labor, while others are increasing investment in AI. Additionally, there could be non-firm-specific factors driving the shift, such as the overall economy or changes in business spending patterns.

We want to identify the direction and magnitude of firm-level substitution. Our identification strategy relies on the hypothesis that firms with a higher share of online labor marketplace spending are more motivated to explore AI tooling. Given the fixed costs of AI tooling exploration, firms are able to amortize the cost of exploration over a larger share of their spend.

We find support for this hypothesis. Firms that spent a higher share of their total spend on online labor marketplaces prior to the ChatGPT introduction spend a higher share of their spend on AI model providers after the introduction of ChatGPT. The effect is stark and linear.

\begin{figure}[ht]
\centering
\includegraphics[width=0.8\textwidth]{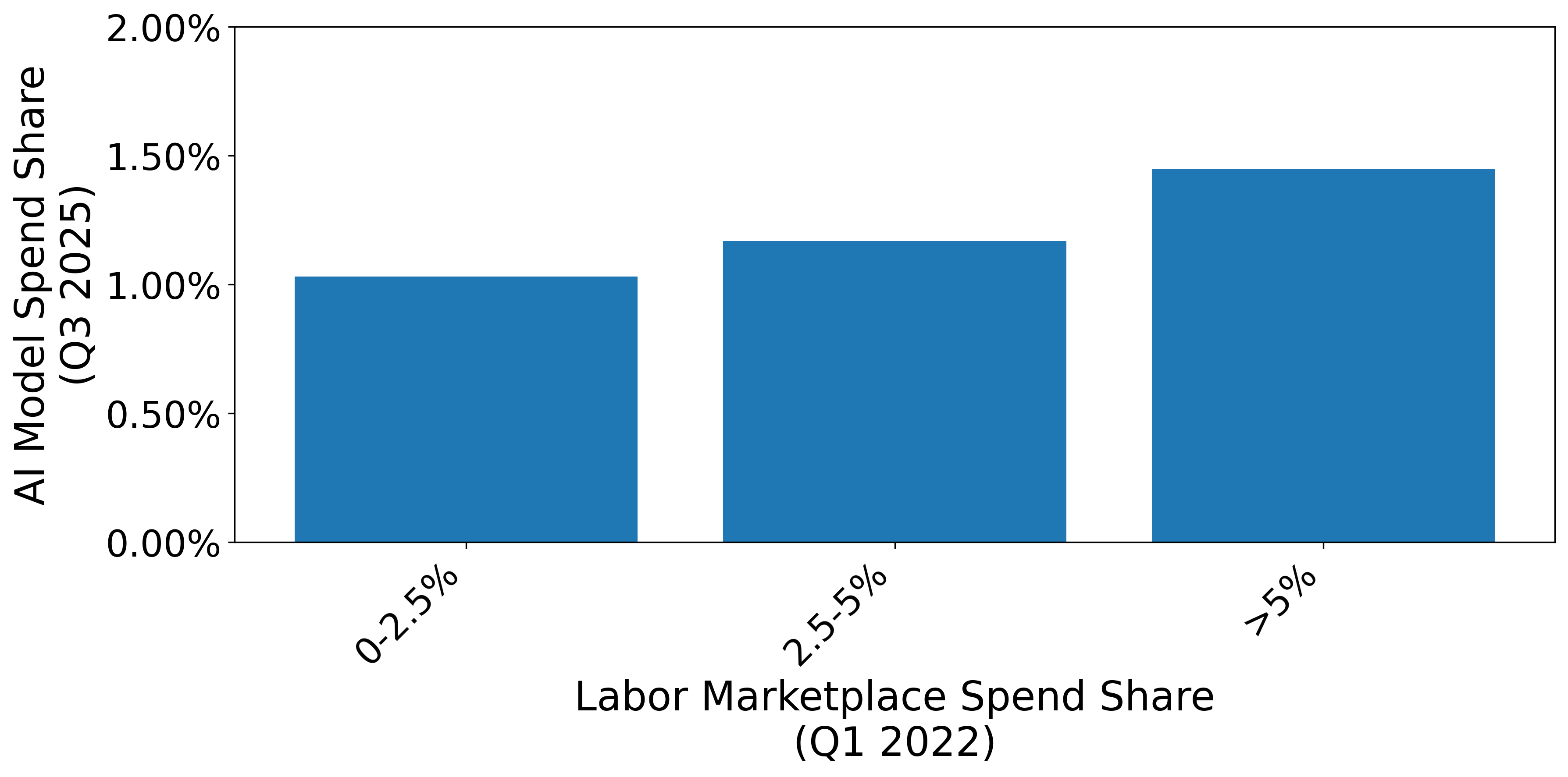}
\caption{Relationship between pre-ChatGPT labor marketplace spending and post-ChatGPT AI model provider spending at the business level}
\label{fig:business_level_relationship}
\end{figure}

In the next section, we will walk through an identification strategy to estimate the causal effect of AI adoption on labor spending.

\section{Identification Strategy}

Our goal is to identify the direction and magnitude of firm-level substitution from labor to AI. We want to know how much AI spending is required to replace a given amount of labor.

We use a difference-in-differences approach to identify the causal effect of AI adoption on labor spending. Similar to Demirci, Hannane, and Zhu (2025), we exploit the natural experiment of the introduction of ChatGPT in October 2022 \cite{DemirciHannaneZhu2025}. ChatGPT led to a sharp increase in the awareness of AI tooling and provides a natural break point to understand how AI adoption impacts other aspects of the economy.

In addition to leveraging the natural experiment, we use variation in the share of spending on online labor marketplaces prior to the introduction of ChatGPT. This share can be viewed as a dosage variable. Firms that were more exposed to potential cost-saving gains from AI tooling will be more motivated to explore and adopt it. This dosage variable allows us to estimate the marginal effect of AI adoption on labor spending. We exclude firms that had no spending on online labor marketplaces in our pre-treatment period because we are not able to measure their exposure to AI shocks.

We use a Two-Way Fixed Effects model to estimate the causal effect of AI adoption on labor spending. We estimate two models, one for the share of spend on online labor marketplaces and one for the share of spend on AI model providers:

\begin{equation}
  \label{eq:main}
  s^{J}_{i, t} = \alpha_{i} + \gamma_{t} + \sum_{k=T_{\text{Q2 2022}}}^{T_{\text{Q3 2025}}} \delta_{k} \cdot \mathbb{I}(k \geq T_{\text{Q3 2022}}) \cdot \mathbb{E}_{i} + \epsilon_{i, t}
\end{equation}

where,
\begin{align*}
  J &= \text{Online Labor Marketplaces (OLM) or AI Model Providers (AI)} \\
  i &= \text{Firm} \\
  t &= \text{Quarter} \\
  s^{J}_{i,t} &= \text{Share of spend on } J \text{ for firm } i \text{ in quarter } t \\
  \alpha_{i} &= \text{Firm fixed effect} \\
  \gamma_{t} &= \text{Quarter fixed effect} \\
  \delta_{k} &= \text{Quarter-specific treatment effect} \\
  \mathbb{E}_{i} &= \text{Exposure to AI-related shocks for firm } i \\
\end{align*}

Our goal is to identify the coefficient on the interaction term, $\delta_{k}$. This coefficient tells us the average change in the share of spend for every unit increase in exposure to AI-related shocks. We compute coefficients for both labor marketplace spend and AI model provider spend. Standard errors are computed using cluster-robust standard errors clustered at the firm level.

To identify exposure to AI-related shocks, we define $\mathbb{E}_{i}$ using the share of spend on online labor marketplaces in Q2 2022, $s^{OLM}_{i, Q2 2022}$. We bucket firms into quartiles of spend; thus, $\mathbb{E}_{i}$ takes on four values. Our hypothesis is that higher quartiles should spend relatively more on AI model providers, as they are more motivated to explore and adopt AI.

\section{Results}

We first present results for the share of spend on AI model providers. We find that firms’ exposure to AI shocks has a significant effect on how much they spend on AI model providers. For the highest quartile of exposure (firms that had $\ge$75\% of their spend on online labor marketplaces in Q2 2022), we find that they significantly increased their share of spending on AI model providers in 2025. In Q3 2025, firms most exposed to AI shocks increased their share of spend on AI by 0.8\% relative to firms least exposed (and relative to Q1–Q2 2022 baselines). Note that this is an absolute 0.8\% share increase, not a relative one. This is fairly significant considering the aggregate share of spend on AI providers is 2.85\% in Q3 2025. On the other hand, for the lowest quartile of exposure (firms that had $\le$25\% of their spend on online labor marketplaces in Q2 2022), we do not find any significant increase. Our results are consistent with the hypothesis that firms more exposed to AI-related shocks are more motivated to explore and adopt AI.

\begin{figure}[ht]
  \centering
  \includegraphics[width=0.8\textwidth]{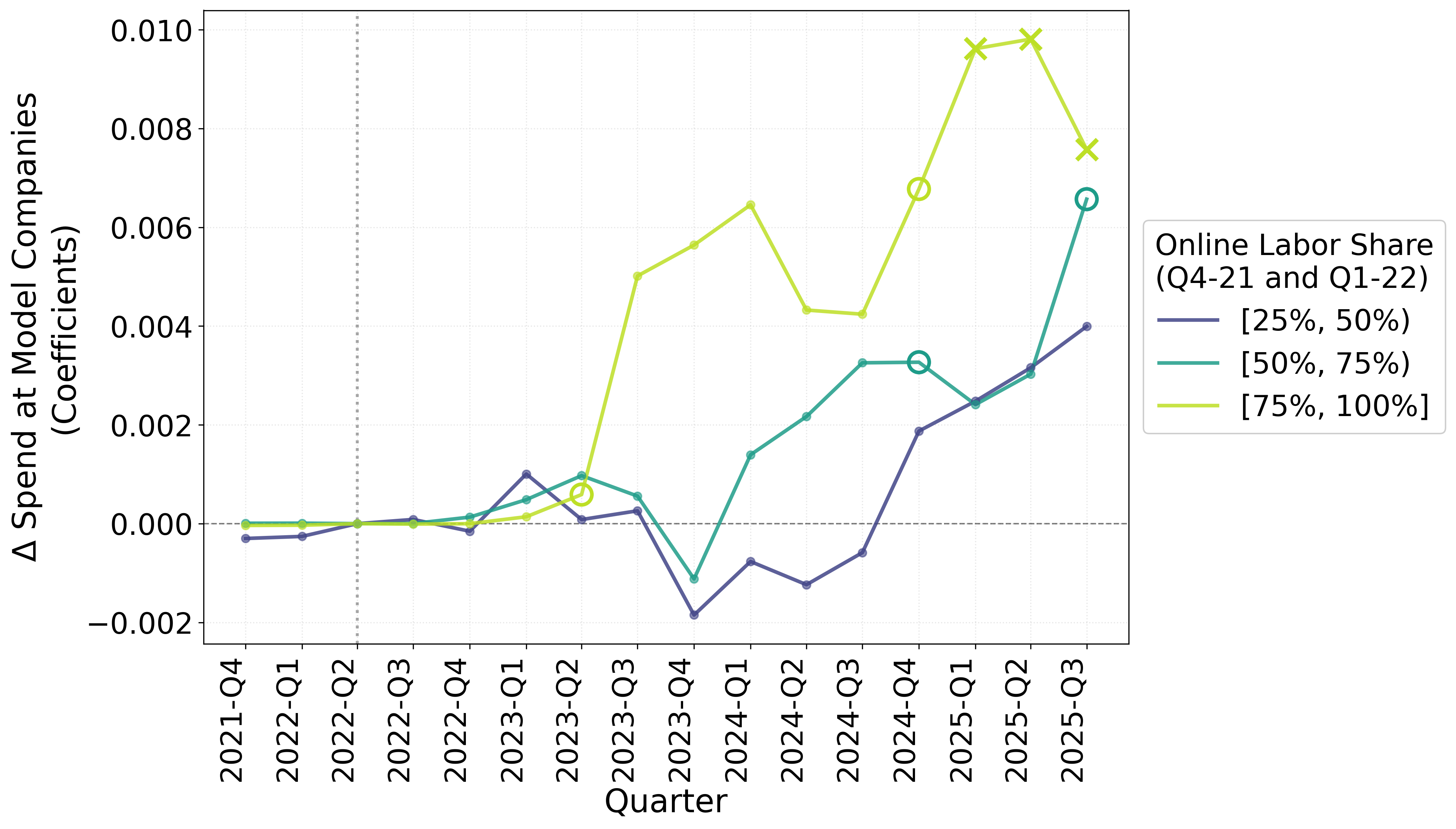}
  \caption{Regression results for AI model provider spending by exposure quartile; X = Significant at 5\% level, O = Significant at 10\% level}
  \label{fig:ai_regression}
  \end{figure}
  
  \clearpage

In addition to the direction and magnitude of AI spending increases, we see differential timing of spending ramp-ups by exposure. Higher-exposed firms ramp up their spending on AI earlier than lower-exposed firms. This is further evidence that higher-exposed firms are more motivated to explore and adopt AI.

Similar to AI model spend, we find that firms most exposed to AI shocks spent relatively less on labor marketplaces than firms least exposed, confirming our hypothesis. The highest-exposed firms spend 15\% less (in absolute terms) on labor marketplaces than firms least exposed. One caveat is that we observe a potential pre-trend issue: the highest-exposed firms had significantly higher spend in Q4 2021, before the AI shock. However, the drop in labor marketplace spend is much larger and persistent over time. Additionally, we find significant decreases in labor marketplace spend among the middle quartile of exposure—those that spent between 50\% and 75\% of their spend on online labor marketplaces in Q2 2022. Their share of spending on labor marketplaces decreased by 2\% in Q3 2025 relative to Q1–Q2 2022 baselines.

\begin{figure}[!htbp]
\centering
\includegraphics[width=0.8\textwidth]{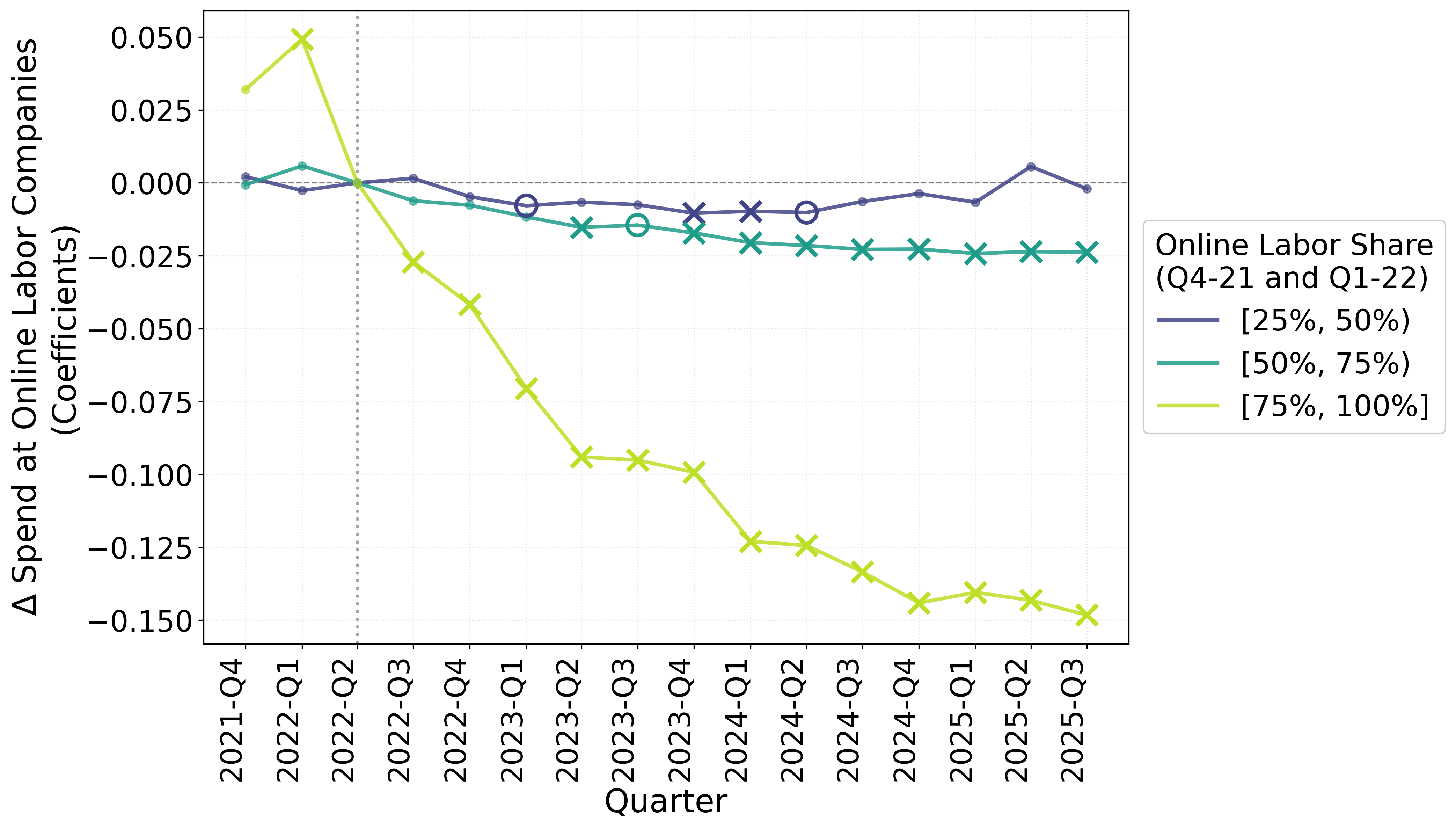}
\caption{Regression results for online labor marketplace spending by exposure quartile; X = Significant at 5\% level, O = Significant at 10\% level}
\label{fig:labor_regression}
\end{figure}

The prior results suggests AI has caused firms to substitute labor for AI. This fits with prior evidence on job posting data from online labor marketplaces.\cite{DemirciHannaneZhu2025} The next question is what is how efficient is AI subsitution for labor? How much increase in AI spend offsets a \$1 reduction in labor marketplace spend is the focus of this section. This is a good proxy for the main question in policy circles: how much is AI replacing labor? To compute how the share of spend shifts from labor marketplaces to AI model providers, we take the ratio of the coefficients from the AI model provider regression and the labor marketplace regression, $\delta_{k}^{AI} / \delta_{k}^{OLM}$. We run a bootstrap simulation ($B = 500$) to compute standard errors for the ratio of the coefficients. Given the wide distribution of ratios for each quartile, we plot each quartile's ratio of coefficients in separate plots. The interpretation of these ratios are straightforward. As an example, assume the ratio is 0.5, this means that for every \$1 decrease in labor marketplace spend, there is a \$0.50 increase in AI model provider spend. We sort the plots below from most exposed businesses to least exposed businesses.

\begin{figure}[H]
\centering
\begin{subfigure}{0.55\textwidth}
\centering
\includegraphics[width=\textwidth]{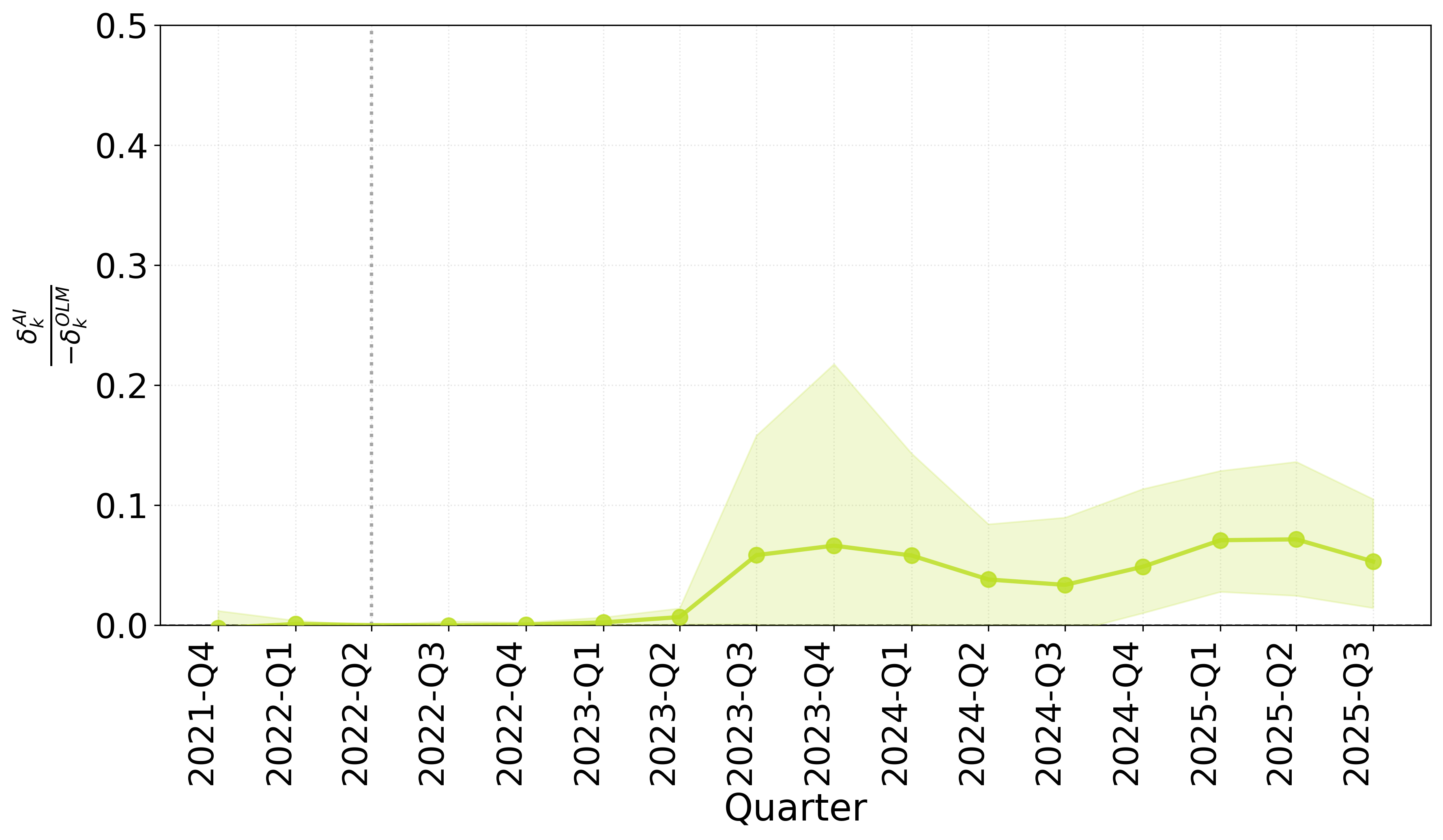}
\caption{Exposure Quartile 3: [75\%, 100\%] labor spend share}
\end{subfigure}
\hfill
\begin{subfigure}{0.55\textwidth}
\centering
\includegraphics[width=\textwidth]{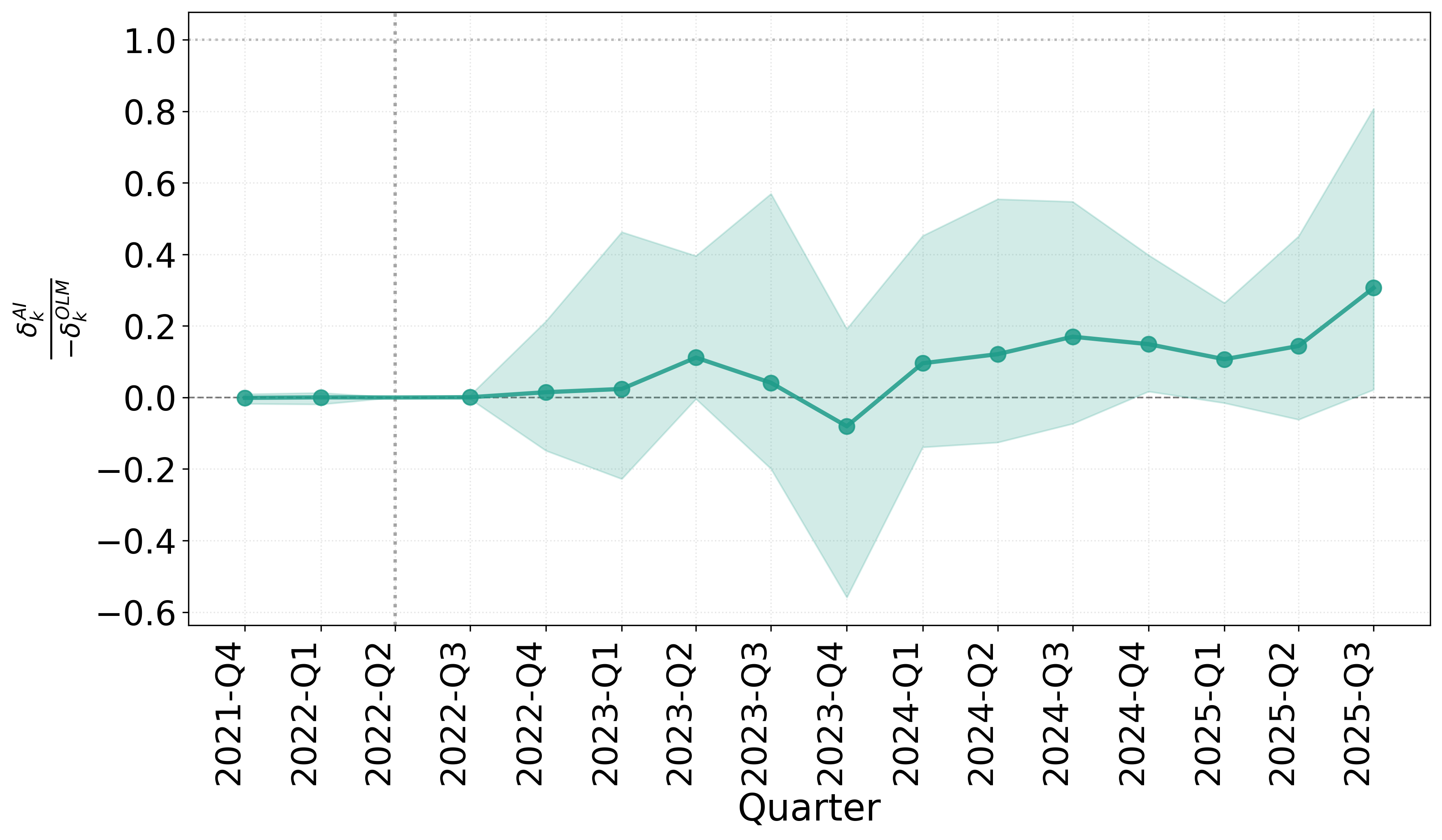}
\caption{Exposure Quartile 2: [50\%, 75\%] labor spend share}
\end{subfigure}

\vspace{0.5cm}

\begin{subfigure}{0.55\textwidth}
\centering
\includegraphics[width=\textwidth]{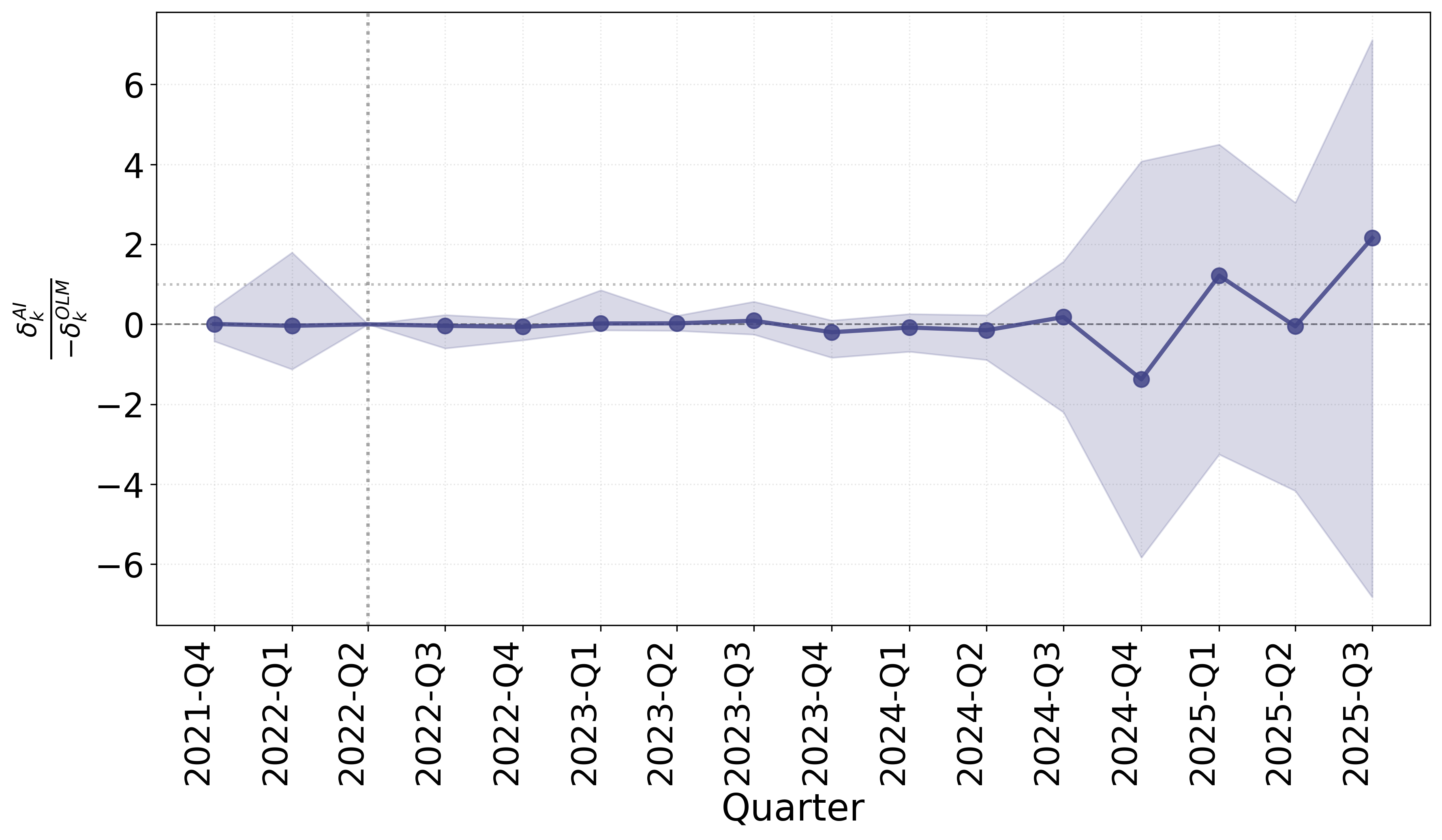}
\caption{Exposure Quartile 1: [25\%, 50\%] labor spend share}
\end{subfigure}

\caption{Bootstrap distribution of coefficient ratios ($\delta_{k}^{AI} / \delta_{k}^{OLM}$) by exposure quartile. Confidence intervals are computed at the 90\% level.}
\label{fig:coef_ratios}
\end{figure}

We see differential patterns of spending shifts by exposure quartile. In the highest exposure quartile, we find that for every \$1 decrease in labor marketplace spend, there is a \$0.03 increase in AI model provider spend in Q3 2025 relative to Q1–Q2 2022 baselines. In the middle exposure quartile, we find that for every \$1 decrease in labor marketplace spend, there is a \$0.30 increase in AI model provider spend in Q3 2025 relative to Q1–Q2 2022 baselines. The true magnitude most likely lies somewhere between these two quartiles. The middle exposure quartile is only significant in the last time period, whereas the highest exposure quartile is significant in all time periods. We note that we cannot observe all potential additional spending that comes from bringing AI in-house, such as infrastructure costs for serving models, as well as increases in engineering headcount to build and maintain AI capabilities.

Even if this estimate is conservative, it is still a significant cost savings. For example, if a firm is spending \$100,000 on labor marketplaces and \$10,000 on AI model providers, the firm is saving \$90,000 by substituting labor for AI. Understanding how these cost savings are distributed both across and within firms is important to understand the potential impact of AI on labor markets and the economy more generally.

\section{Conclusion}

This study sheds light on how businesses are substituting labor for AI; however, the magnitude of the substitution is not uniform. Increasing exposure to AI shocks causes businesses to substitute away from labor marketplaces faster and at a higher rate. To our knowledge, this is the first study to show firm-level differences in substitution patterns.

Substitution from labor to AI is not simply faster but also appears to be happening at a lower cost. Higher AI-exposed firms substitute relatively more labor for AI at a lower cost. While this study provides evidence, we cannot determine the mechanism behind the substitution. There are many different theories that could explain this pattern, such as returns to scale in building internal AI capabilities or having a natural advantage in AI adoption relative to less exposed firms. Exploring these mechanisms is beyond the scope of this study but is an important area for future research.

\printbibliography

\end{document}